\documentclass[twocolumn,showpacs,preprintnumbers, prl,
               superscriptaddress]{revtex4}
\usepackage{graphicx, fancybox}
\usepackage{amsmath,amssymb}

\begin{document}


\title{
Revealing baryon number fluctuations from proton number fluctuations 
\\ 
in relativistic heavy ion collisions 
}

\author{Masakiyo Kitazawa}
\email{kitazawa@phys.sci.osaka-u.ac.jp}
\affiliation{
Department of Physics, Osaka University, Toyonaka, Osaka 560-0043, Japan}

\author{Masayuki Asakawa}
\email{yuki@phys.sci.osaka-u.ac.jp}
\affiliation{
Department of Physics, Osaka University, Toyonaka, Osaka 560-0043, Japan}

\begin{abstract}

Baryon number cumulants are invaluable 
tools to diagnose the primordial stage of 
heavy ion collisions if they can be measured.
In experiments, however, proton number cumulants have been
measured as substitutes.
In fact, proton number fluctuations are further modified 
in the hadron phase and are different from those of 
the baryon number.
We show that the isospin distribution of 
nucleons at kinetic freeze-out is binomial and
factorized. This leads to formulas that
express the baryon number cumulants
solely in terms of proton number fluctuations,
which are experimentally observable.
\end{abstract}

\date{\today}

\maketitle

The order of the phase transition of quantum chromodynamics 
(QCD) at nonzero temperature ($T$) is believed to change 
from crossover \cite{lattice} to first order 
at a nonzero baryon chemical potential ($\mu_{\rm B}$).
The existence of the QCD critical point is thus expected 
in the phase diagram on the $T$-$\mu_{\rm B}$ plane
\cite{Stephanov:2004wx}.
Experiments to explore the phase structure 
at nonzero $\mu_{\rm B}$, especially the 
existence of the critical point, are now ongoing in the energy 
scan program at the Relativistic Heavy Ion Collider (RHIC) 
\cite{STAR,Mohanty:2011nm}, and 
will also be performed in future facilities \cite{FAIR,NICA}.
Much attention has also been paid to this problem from 
numerical experiments on the lattice \cite{Gavai:2010zn,lattice}.
The establishment of the QCD phase structure at nonzero $\mu_{\rm B}$
is an important issue, not only to 
deepen our knowledge of the matter described by QCD, 
but also to gain understanding of a wide array of topics 
in physics which share the concepts of 
phase transitions and techniques to treat 
strongly correlated many-body systems.

Fluctuations, which are experimentally measured by event-by-event 
analyses in heavy ion collisions, are promising observables 
to probe the properties of created fireballs \cite{Koch:2008ia}, 
as their behaviors are sensitive to the state of the matter.
For example, because of the singularity at the critical 
point, fluctuations of various physical quantities, including skewness 
and kurtosis, behave anomalously near the critical point 
\cite{Stephanov:1998dy,Hatta:2003wn,Stephanov:2008qz}.
One can also argue that ratios between the cumulants of conserved
charges are sensitive to the magnitudes of the charge carried 
by the quasiparticles composing the system, and hence they 
behave differently in the hadronic and quark-gluon phases
\cite{Asakawa:2000wh,Jeon:2000wg,Ejiri:2005wq}.
Recently, it was also pointed out that some higher-order 
cumulants of conserved charges change signs around the phase 
boundary of QCD, which would serve as clear experimental 
signatures to determine the location of the matter in the phase 
diagram \cite{Asakawa:2009aj,Friman:2011pf,Stephanov:2011pb}.

Among the fluctuation observables, those of conserved charges
can reflect fluctuations produced in earlier stages
during the time evolution of fireballs, than non-conserved ones
\cite{Stephanov:2009ra}.
This is because the variation of a conserved charge in a volume 
is achieved only through diffusion, which makes the relaxation 
to equilibrium slower.
In fact, it is argued that if the rapidity range of a detector 
is taken to be sufficiently large, 
whereas the range should be kept narrow enough so that 
the rest of fireballs can be regarded as the heat bath, 
the effects of diffusion 
are well suppressed and fluctuations produced in the 
quark-gluon phase can be detected experimentally
\cite{Asakawa:2000wh,Jeon:2000wg}.

The dependences of the proton number fluctuations, 
cumulants up to fourth order, on the beam energy $\sqrt{s}$, 
have been recently measured by the STAR collaboration at RHIC 
\cite{STAR,Mohanty:2011nm}.
The result appears to be almost consistent with the prediction of the 
hadron resonance gas (HRG) model \cite{Karsch:2010ck}; although 
the experimental result shows some deviation from the prediction 
at small $\sqrt{s}$, it is at most of the order of $20\%$ 
\cite{Mohanty:2011nm}. 
The proton number, however, is not a conserved quantity, 
and in fact we will see later that its fluctuations 
significantly evolve in the hadronic stage, which makes the 
experimentally measured fluctuations close to those in the 
equilibrated hadronic matter.
The agreement between the experiments and the HRG model in 
the proton number fluctuations \cite{Mohanty:2011nm} is in part 
due to these effects, and hence it does not immediately exclude 
the slow baryon number diffusion in the hadronic stage.
Although the measurement of the baryon number, which is a conserved
charge, is desirable to probe fluctuations generated in earlier 
stages, its direct experimental measurement has been considered
to be impossible because of the difficulty in 
detecting and identifying neutrons.

In this Rapid Communication, we show that the experimentally measured 
proton number fluctuations are nevertheless directly related
to baryon number fluctuations in earlier stages, and 
we present concrete formulas that relate the baryon
number cumulants and these experimental observables.
The key observation is that 
the distributions of (anti-)proton and (anti-)neutron numbers 
in the final state are well described by binomial distributions.
As will be argued in detail later, this observation 
is well justified at least for RHIC energy, and is expected
to hold for $\sqrt{s}\gtrsim10{\rm GeV}$.

Experimentally, the electric charge can be measured directly.
Electric charge fluctuations, however, contain the contribution of 
isospin fluctuations, which are non-singular at the critical point, 
in addition to baryon number fluctuations \cite{Hatta:2003wn}.
The signals of the phase transition in this observable thus 
generally become weak owing to the non-singular contribution
(such a tendency, for example, in the third moments is seen 
in Ref.~\cite{Asakawa:2009aj}).
In this sense, the baryon number fluctuations are superior to
the electric ones as probes of the QCD phase structure.

Throughout this Rapid Communication, we use $N_X$ to represent 
the number of particles $X$ leaving the system after each 
collision event, where $X=p$, $n$, and B 
represent proton, neutron, and baryon, respectively, and 
their anti-particles, $\bar p$, $\bar n$, and $\bar{\rm B}$.
The net and total numbers are denoted as 
$N_X^{\rm (net)} = N_X - N_{\bar{X}}$ and 
$N_X^{\rm (tot)} = N_X + N_{\bar{X}}$, respectively.

Before starting the main discussion on the cumulants of baryon and 
proton numbers, let us briefly consider how the proton 
number fluctuations evolve in the hadronic stage.
The most important process responsible for the variation of the 
proton number is the charge exchange reactions with thermal pions
mediated by $\Delta^+(1232)$ and $\Delta^0(1232)$ resonances:
\begin{align}
p(n) + \pi \to \Delta^{+,0} \to n(p) + \pi.
\label{eq:reaction}
\end{align}
Because of the small energy required and the large cross 
sections, these reactions proceed even after chemical
freeze-out, as we demonstrate later.
We note that these reactions do not alter the average abundances
$\langle N_p \rangle$ and $\langle N_{\bar p} \rangle$ 
if the isospin chemical potential vanishes,
while they modify the fluctuations of $N_p$ and $N_{\bar p}$.
Because chemical freeze-out is a concept that describes 
ratios between particle abundances such as 
$\langle N_{\bar p} \rangle / \langle N_p \rangle$,
these reactions below the chemical freeze-out temperature 
$T_{\rm chem}$ do not contradict the statistical model.
The success of the model, on the other hand, indicates that 
the creation and annihilation of (anti-)nucleons hardly occur 
below $T_{\rm chem}$.

The importance of reactions (\ref{eq:reaction}) below 
$T_{\rm chem}$ is confirmed by evaluating 
the mean time of the nucleons for these reactions.
Provided that the pions have a thermal distribution, the mean time 
$\tau$ that a proton at rest in the medium forms $\Delta^+$ or 
$\Delta^0$, being scattered by a thermal pion, is evaluated to be
\begin{align}
\tau^{-1} = \int \frac{d^3 k_\pi}{(2\pi)^3} \sigma(E_{\rm c.m.}) v_\pi n(E_\pi),
\label{eq:tau}
\end{align}
with the Bose distribution function $n(E)= ( e^{E/T}-1)^{-1}$,
pion velocity $v_\pi=k_\pi/E_\pi$, $E_\pi=\sqrt{m_\pi^2+k_\pi^2}$, 
and the pion mass $m_\pi$.
$\sigma(E_{\rm c.m.})$ is the sum of the cross sections for $p\pi$ 
reactions producing $\Delta^+$ and $\Delta^0$ with a 
center-of-mass energy $E_{\rm c.m.}=[ (m_{\rm N}+E_\pi)^2 - k_\pi^2]^{1/2}$ 
with the nucleon mass $m_{\rm N}$.
To evaluate Eq.~(\ref{eq:tau}), we assume a cross section of 
Breit-Wigner type, $\sigma(E_{\rm c.m.}) =
\sigma_\Delta (\Gamma^2/4) / ( (E_{\rm c.m.}-E_\Delta)^2 + \Gamma^2/4 )$,
which is a sufficient approximation for our purpose,
with hadron properties in the vacuum, $m_N=940{\rm MeV}$, 
$m_\pi=140{\rm MeV}$, $E_\Delta=1232{\rm MeV}$, $\Gamma=110{\rm MeV}$, 
and $\sigma_\Delta=20{\rm fm}^2$ \cite{PDG}.
The mean time is then evaluated to be $3-4$ fm for 
$T=150-170{\rm MeV}$. 
One can also check that this mean time hardly changes even for 
moving protons in the range of momentum $p\lesssim 3T$.
On the other hand, dynamical models for RHIC energy predict 
that protons stay in the hadronic gas and continue to interact
for several tens of fm 
on average at midrapidity \cite{Nonaka:2006yn}, which is 
significantly longer than the mean time and the lifetime of 
$\Delta$, $1/\Gamma\simeq1.8{\rm fm}$.
This result shows that nucleons in the fireball indeed undergo 
this reaction several times on average in the hadronic stage
\footnote{
We note that some event generators employed in 
Ref.~\cite{STAR} do not take this reaction into account. 
As we will see later, however, it is this reaction that is 
responsible for our main results Eq.~(\ref{eq:NB}) - (\ref{eq:NB4}) 
expressing the baryon number fluctuations in terms of experimental 
observables without introducing any models. 
}.
The ratio of the probabilities that a proton in the medium produces 
a $\Delta^+$ or $\Delta^0$ and then decays into $p$ and $n$ 
is $5:4$, which is determined by the isospin SU(2) algebra.
Whereas this probability is not even, after repeating
the above processes several times
in the hadronic stage, the nucleons tend to completely 
forget their initial isospin.

The above discussion shows that the evolution of proton number 
fluctuations in the hadronic stage is dominantly made via the 
exchanges of the two isospin states of the nucleons.
Now, we further assert that {\it isospins of all nucleons 
in the final state are uncorrelated}.
This statement is well justified when the hadronic medium 
fulfills the following two conditions: (i) The medium effects on the 
branching ratios and formation rates of $\Delta$ are insensitive 
to the proton and neutron number densities $n_p$ and $n_n$
(and the same holds for the anti-particle sector as well), 
and (ii) (anti-)nucleon-(anti-)nucleon 
interactions generating correlations between two nucleons 
hardly occur.
As we will see later, these two conditions are well satisfied 
below $T_{\rm chem}$ except for low-energy collisions.
The probability distribution of finding $N_p$ and $N_n$ 
($N_{\bar p}$ and $N_{\bar n}$) 
particles in the final state 
in each event then becomes binomial.
Under the isospin symmetry \cite{Karsch:2010ck},
this fact enables to factorize the 
probability distribution $P( N_p , N_n , N_{\bar p} , N_{\bar n} )$ 
having $N_p$, $N_n$, $N_{\bar p}$, and $N_{\bar n}$ particles 
in each event as 
\begin{align}
\lefteqn{P( N_p , N_n , N_{\bar p} , N_{\bar n} ) }
\nonumber \\
&= F( N_{\rm B} , N_{\bar{\rm B}} ) 
B( N_p ; N_{\rm B} ) B( N_{\bar p} ; N_{\bar{\rm B}} ),
\label{eq:P}
\end{align}
where $B(k;N)= 2^{-N} N!/(k!(N-k)!)$ is the binomial distribution
function with an equal probability.
On the right-hand side (RHS) of Eq.~(\ref{eq:P}) we have used 
$N_{\rm B}$ and $N_{\bar{\rm B}}$ defined by
$N_{\rm B} = N_p+N_n$ and
$N_{\bar{\rm B}} = N_{\bar p}+N_{\bar n}$.
We will later elucidate this notation to use the baryon 
numbers $N_{\rm B}$ and $N_{\bar{\rm B}}$ in place of the 
nucleon numbers.
Under the probability distribution Eq.~(\ref{eq:P}),
the event-by-event average of a function
$f( N_p , N_{\bar p} )$ is given by 
\begin{align}
\left\langle f( N_p , N_{\bar p} )
\right\rangle
=& \sum_{N_{\{ p,n,{\bar p},{\bar n}\}}}
P(N_p,N_n,N_{\bar p},N_{\bar n}) f( N_p , N_{\bar p} )
\nonumber 
\\
=& \sum_{N_{\rm B},N_{\bar{\rm B}}} F( N_{\rm B} , N_{\bar{\rm B}} ) 
\sum_{N_p,N_{\bar p}} f( N_p , N_{\bar p} )
\nonumber 
\\
&\times 
B( N_p ; N_{\rm B} ) B( N_{\bar p} ; N_{\bar{\rm B}} ).
\label{eq:average}
\end{align}

The factorization Eq.~(\ref{eq:P}) leads to 
\begin{align}
\langle N_p^{\rm (net)} \rangle
=& \frac12 \langle N_{\rm B}^{\rm (net)} \rangle,
\label{eq:Np}
\\
\langle (\delta N_p^{\rm (net)})^2 \rangle
=& \frac14 \langle (\delta N_{\rm B}^{\rm (net)})^2 \rangle
+ \frac14 \langle N_{\rm B}^{\rm (tot)} \rangle,
\label{eq:Np2}
\\
\langle (\delta N_p^{\rm (net)})^3 \rangle
=& \frac18 \langle (\delta N_{\rm B}^{\rm (net)})^3 \rangle
+ \frac38 \langle \delta N_{\rm B}^{\rm (net)} \delta N_{\rm B}^{\rm (tot)} \rangle,
\label{eq:Np3}
\\
\langle (\delta N_p^{\rm (net)})^4 \rangle_c
\equiv& \langle (\delta N_p^{\rm (net)})^4 \rangle 
- 3 \langle (\delta N_p^{\rm (net)})^2 \rangle^2 
\nonumber \\
=& \frac1{16} \langle (\delta N_{\rm B}^{\rm (net)})^4 \rangle_c
+ \frac38 \langle (\delta N_{\rm B}^{\rm (net)})^2 \delta N_{\rm B}^{\rm (tot)} \rangle
\nonumber \\
&
+ \frac3{16} \langle (\delta N_{\rm B}^{\rm (tot)})^2 \rangle
- \frac18 \langle N_{\rm B}^{\rm (tot)} \rangle,
\label{eq:Np4}
\end{align}
where $\delta N_X = N_X - \langle N_X \rangle$.
To derive Eqs.~(\ref{eq:Np})$-$(\ref{eq:Np4}),
we have used 
the fact that the sums over $N_p$ and $N_{\bar p}$ in 
Eq.~(\ref{eq:average}) can be taken separately with corresponding 
binomial functions, {\it e.g.}, 
$\sum_{N_p} N_p B( N_p; N_{\rm B} ) = N_{\rm B}/2$ and 
$\sum_{N_p} N_p^2 B( N_p; N_{\rm B} ) = N_{\rm B}^2/4 + N_{\rm B}/4$.

Equation~(\ref{eq:P}) also enables to represent the baryon number 
cumulants by those of the net and total proton numbers as
\begin{align}
\langle N_{\rm B}^{\rm (net)} \rangle
=& 2 \langle N_p^{\rm (net)} \rangle ,
\label{eq:NB}
\\
\langle (\delta N_{\rm B}^{\rm (net)})^2 \rangle
=& 4 \langle (\delta N_p^{\rm (net)})^2 \rangle
-2 \langle N_p^{\rm (tot)} \rangle ,
\label{eq:NB2}
\\
\langle (\delta N_{\rm B}^{\rm (net)})^3 \rangle
=& 8 \langle (\delta N_p^{\rm (net)})^3 \rangle
-12 \langle \delta N_p^{\rm (net)} \delta N_p^{\rm (tot)} \rangle
\nonumber \\
&+6 \langle N_p^{\rm (net)} \rangle ,
\label{eq:NB3}
\\
\langle (\delta N_{\rm B}^{\rm (net)})^4 \rangle_c
=& 16 \langle (\delta N_p^{\rm (net)})^4 \rangle_c
-48 \langle (\delta N_p^{\rm (net)})^2 \delta N_p^{\rm (tot)} \rangle
\nonumber \\
&
+ 48 \langle (\delta N_p^{\rm (net)})^2 \rangle
+ 12 \langle (\delta N_p^{\rm (tot)})^2 \rangle
\nonumber \\
& 
- 26 \langle N_p^{\rm (tot)} \rangle ,
\label{eq:NB4}
\end{align}
where we have used relations for mixed cumulants such as 
$\langle \delta N_{\rm B}^{\rm (net)} \delta N_{\rm B}^{\rm (tot)} \rangle
= 4 \langle \delta N_p^{\rm (net)} \delta N_p^{\rm (tot)} \rangle
- 2 \langle N_p^{\rm (tot)} \rangle$
which are obtained with Eq.~(\ref{eq:P}).
Since the RHSs of Eqs.~(\ref{eq:NB}) - 
(\ref{eq:NB4}) consist of only $N_p^{\rm (net)}$ and $N_p^{\rm (tot)}$,
which are experimentally observable, these are formulas that
express baryon number cumulants solely in terms of experimental 
observables.
We remind that no specific form of $F(N_{\rm B},N_{\bar{\rm B}})$ 
is assumed in deriving these results.

We remark that $N_{\rm B}^{\rm (net)}$ ($N_{\rm B}^{\rm (tot)}$) 
in Eqs.~(\ref{eq:Np}) - (\ref{eq:NB4}) are interpreted to be 
the sum of all net (total) baryon numbers entering 
a region in the phase space in the final state of each event.
If the diffusion of the baryon number in the hadronic stage is slow 
\cite{Asakawa:2000wh,Jeon:2000wg}, 
the information on the primordial fluctuations remains
in $F(N_{\rm B},N_{\bar{\rm B}})$ in Eq.~(\ref{eq:P}) 
and, as a result, in baryon number cumulants.

Next, let us inspect the validity of Eq.~(\ref{eq:P}) in more detail.
First, we consider the conditions (i) and (ii) introduced above 
Eq.~(\ref{eq:P}).
In the medium, the decay rate of $\Delta$ acquires the statistical factor 
\begin{align}
\left( 1-f(E_N) \right) \left( 1+n(E_\pi) \right),
\label{eq:f}
\end{align}
where $f(E)=(e^{(E-\mu_B)/T}+1)^{-1}$ is the Fermi distribution function 
and $E_N$ and $E_\pi$ are the energies of the nucleon and pion produced 
by the decay, respectively.
The first term in Eq.~(\ref{eq:f}) represents the 
Pauli blocking effect.
At RHIC energy, the Boltzmann approximation is well applied to nucleons 
below $T_{\rm chem}$ since $T\ll m_N$ and $|\mu_{\rm B}|\ll m_N$.
Thus, the Pauli blocking effect can be almost ignored.
The Bose factor $(1+n(E_\pi))$ in Eq.~(\ref{eq:f}), on the other
hand, has a nonnegligible contribution since 
$m_\pi\simeq T_{\rm chem}$.
The density of the pions, however, is more than one order larger than 
that of the nucleons below $T_{\rm chem}$.
The Bose factor thus must be insensitive to 
$n_p$ and $n_n$, while it leads to the enhancement of the decay of 
$\Delta$ in the medium, which acts in favor of the isospin
randomization. 
The large pion density also means that the mean time for a 
nucleon to form $\Delta$ is insensitive to $n_p$ and $n_n$.
Condition (i) is thus well satisfied 
below $T_{\rm chem}$ at RHIC energy.
The validity of condition (ii) is conjectured from 
the success of the statistical model as follows.
The statistical model indicates that the pair annihilation 
of a N and an $\bar{\rm N}$ terminates at $T_{\rm chem}$.
NN and N$\bar{\rm N}$ reactions are then also expected 
to terminate there, because the elastic cross section of 
N$\bar{\rm N}$ is significantly smaller than the inelastic one, 
and the total cross section of NN behaves similarly to that 
of N$\bar{\rm N}$ for $E_{\rm c.m.}<1$GeV \cite{PDG}.
Condition (ii) thus should also be satisfied for $T<T_{\rm chem}$.
Intuitively speaking, in a hot medium the nucleons are 
so dilutely distributed that they do not feel one another's existence,
while there are so many pions which can be regarded as 
the heat bath when the nucleon sector is concerned.
The large pion density also enables to use the binomial 
distribution independently of the initial nucleon isospin 
density.

Second, while so far we have limited our attention to the 
nucleon reactions mediated by $\Delta$, other 
interactions can also take place in the medium.
It is also possible that $\Delta$ interacts with a 
thermal pion to form another resonance before the decay 
\cite{Pang:1992sk}.
All these reactions with thermal pions, however, proceed 
with a certain probability determined by the isospin SU(2) 
symmetry as long as they are caused by the strong interaction, 
and the reactions of a baryon make its isospin random.
Strange baryons, on the other hand, decay via the weak or 
electromagnetic interaction outside the fireball. 
In particular, $\Lambda$ and $\Sigma$ are important among them.
$\Lambda$ decays into $p$ and $n$ with a branching ratio of $16:9$. 
Provided that the three isospin states of $\Sigma$ are produced with an 
equal probability in the medium, the ratio of probabilities 
that a $\Sigma$ decays into $p$ and $n$ is about $1:1.6$
\cite{PDG}.
Although these ratios are not even, because the abundances of 
$\Lambda$ and $\Sigma$ are small compared to the nucleons, to a 
first approximation it is suitable for our purpose to regard 
these probabilities to be equal and to incorporate nucleons produced 
by the decays of $\Lambda$ and $\Sigma$ in $N_p$ and $N_n$ in 
Eq.~(\ref{eq:P}).
This promotes the nucleon numbers
to those of the baryons in Eq.~(\ref{eq:P}).
The treatment of strange baryons, however, may require more 
detailed arguments, especially on their quantitative effects on 
higher-order cumulants, which will be addressed elsewhere.
Inclusion of higher baryonic resonances and light nuclei
such as deuterons will not affect our conclusions
owing to their negligible abundances.

While the factorization Eq.~(\ref{eq:P}) is fully established 
for RHIC energy, the binomiality will eventually break down
as the beam energy is decreased.
At very low beam energy, pions are not produced
enough and nucleons will not undergo
charge exchange reactions sufficiently below $T_{\rm chem}$.
We deduce that this happens when $T_{\rm chem} \lesssim m_\pi$.    
When the reactions hardly occur, the isospin correlations 
generated at the hadronization will remain until the final state.
At low beam energy, also the nucleon density becomes comparable 
with that of the pions, and the latter can no longer be regarded as 
the heat bath to absorb the isospin fluctuations of the former.
From the $\sqrt{s}$ dependence of the chemical freeze-out line on
the $T$-$\mu_{\rm B}$ plane \cite{Cleymans:1998fq}, and considering
the validity of these two conditions, we deduce that 
Eq.~(\ref{eq:P}) is well applicable to the range of beam energy 
$\sqrt{s} \gtrsim 10$~GeV.

In the argument to derive Eqs.~(\ref{eq:Np})$-$(\ref{eq:NB4}), 
we have implicitly assumed that the hadronic 
medium is isospin symmetric.
While the effect of nonzero isospin density should be 
well suppressed for large $\sqrt{s}$ where a large number of 
particles having nonzero isospin charges are produced,
at lower energies this effect gives rise to a non-negligible
modification of Eqs.~(\ref{eq:Np})$-$(\ref{eq:NB4}).
When the system has nonzero isospin density, the probability
that a nucleon at the early stage of the hadron phase becomes 
a proton or a neutron in the final state is no longer even.
This effect is, as long as conditions (i) and (ii) 
introduced above Eq.~(\ref{eq:P}) hold, 
incorporated into our results by 
simply replacing the binomial function $B(N_p;N_{\rm B})$ in 
Eq.~(\ref{eq:P}) with that having a probability 
$k=\langle N_p \rangle/\langle N_p + N_n \rangle$,
and a similar replacement to $B(N_{\bar p};N_{\bar{\rm B}})$.
Our explicit analysis indicates that the effect of nonzero 
isospin density on Eqs.~(\ref{eq:Np})$-$(\ref{eq:NB4}) 
is relatively small and well suppressed when 
$T_{\rm chem} > m_\pi$ and a sufficient number of pions having 
isospin charges are produced at chemical freeze-out.
Since this modification requires a straightforward but lengthy
calculation, we will elucidate the analysis in
a forthcoming paper.

Now, let us apply our results to the latest experimental 
data from STAR \cite{STAR,Mohanty:2011nm}.
To estimate how the binomial nature of nucleon isospins affects
the proton number fluctuations, we first consider 
Eqs.~(\ref{eq:Np})$-$(\ref{eq:Np4}).
In order to estimate the contributions of terms including 
$N_{\rm B}^{\rm (tot)}$ in these equations, 
we temporarily postulate that $N_{\rm B}$ and $N_{\bar{\rm B}}$ 
have thermal distributions fixed at chemical freeze-out
as the statistical model suggests, while the distribution of 
their combination, $N_{\rm B}^{\rm (net)}$, deviates from
the thermal one reflecting the baryon number conservation.
Under this assumption, the distributions of $N_{\rm B}$ and 
$N_p$ are Poissonian, and hence 
the cumulants of the baryon and proton numbers satisfy
\begin{align}
\lefteqn{\langle N_{\rm B} \rangle
= \langle (\delta N_{\rm B})^2 \rangle
= \langle (\delta N_{\rm B})^3 \rangle}
\nonumber \\
&= 2 \langle N_p \rangle_{\rm HG} 
= 2 \langle ( \delta N_p )^2 \rangle_{\rm HG}
= 2 \langle ( \delta N_p )^3 \rangle_{\rm HG},
\end{align}
and the same for anti-baryon numbers, where 
$\langle \cdot \rangle_{\rm HG}$ is the expectation value for 
free hadron gas (HG) composed of mesons and nucleons at 
$T_{\rm chem}$, {\it i.e.} a simplified version of the HRG 
model \cite{Karsch:2010ck}.
Equations~(\ref{eq:Np2}) and (\ref{eq:Np3}) are then expressed as
\begin{align}
\langle (\delta N_p^{\rm (net)} )^2 \rangle
&= \frac14 \langle (\delta N_{\rm B}^{\rm (net)})^2 \rangle
+ \frac12 \langle (\delta N_p^{\rm (net)} )^2 \rangle_{\rm HG},
\label{eq:Np2:HG}
\\
\langle (\delta N_p^{\rm (net)} )^3 \rangle
&= \frac18 \langle (\delta N_{\rm B}^{\rm (net)})^3 \rangle
+ \frac34 \langle (\delta N_p^{\rm (net)} )^3 \rangle_{\rm HG}.
\label{eq:Np3:HG}
\end{align}
To derive these results, we decomposed, for example, 
the second term in Eq.~(\ref{eq:Np3}) as 
\begin{align}
&\langle \delta N_{\rm B}^{\rm (net)} \delta N_{\rm B}^{\rm (tot)} \rangle 
= \langle (\delta N_{\rm B})^2 \rangle 
- \langle (\delta N_{\bar{\rm B}})^2 \rangle
\nonumber \\
&= 2 \langle ( \delta N_p )^3 \rangle_{\rm HG}
- 2 \langle ( \delta N_{\bar p} )^3 \rangle_{\rm HG}
= 2\langle ( \delta N_p^{\rm (net)} )^3 \rangle_{\rm HG}, 
\end{align}
The results in Eqs.~(\ref{eq:Np2:HG}) and (\ref{eq:Np3:HG})
show that the second terms on the RHSs, which come
from the binomial distributions of nucleon isospin, make a large 
contribution to the cumulants of the proton number, and they
become more significant as the order increases.
Although one cannot derive a similar result for 
the fourth-order relation, from the factor $1/16$ in the 
first term of Eq.~(\ref{eq:Np4}) it is clear that the 
effect of the fourth-order baryon number cumulant on the 
proton number one is more suppressed in this order.
The suppression of the first term
in Eqs.~(\ref{eq:Np2})$-$(\ref{eq:Np4}) may be one of 
the reasons why the results of the STAR experiment and the 
HRG model appear to be consistent with each other.
In this sense, it is interesting that the experimental 
results for skewness and kurtosis have small but significant 
deviations from the HRG predictions at $\sqrt{s}\lesssim50$~GeV
\cite{Mohanty:2011nm}.
The deviation, for example, in skewness, can be a consequence 
of $\langle (\delta N_{\rm B}^{\rm (net)})^3 \rangle$ in 
Eq.~(\ref{eq:Np3:HG}), which possibly reflects
the properties of matter in the early stage.
Baryon number cumulants are, of course, directly determined
with experimental observables using Eqs.~(\ref{eq:NB})$-$(\ref{eq:NB4}).
It is worth emphasizing that the RHSs of Eqs.~(\ref{eq:NB3}) and 
(\ref{eq:NB4}) have terms which would lead to the negative 
cumulants discussed in 
Refs.~\cite{Asakawa:2009aj,Stephanov:2011pb},
or the suppression of the ratio 
$\langle (\delta N_{\rm B}^{\rm (net)})^4 \rangle /
\langle (\delta N_{\rm B}^{\rm (net)})^2 \rangle$ 
\cite{Ejiri:2005wq}.

We note that when the distribution of $N_{\rm B}^{\rm (net)}$ 
also follows that in the HG, in addition to the above postulation, 
the RHSs of Eqs.~(\ref{eq:Np2:HG}) and (\ref{eq:Np3:HG}) reduce 
to $ \langle (\delta N_p^{\rm (net)} )^2 \rangle_{\rm HG}$ and 
$ \langle (\delta N_p^{\rm (net)} )^3 \rangle_{\rm HG}$.
A way to check this is to use the fact that the 
(anti-)nucleon numbers in the HG are well described by 
the Poisson distribution owing to the Boltzmann approximation, 
and that the Poisson distribution with an average $\lambda$, 
$P_\lambda(N)$, satisfies
$
P_\lambda(N_1) P_\lambda(N_2) = P_{2\lambda}( N_1+N_2 ) B( N_1 ; N_1+N_2 ).
$
The HG thus corresponds to a special case of Eq.~(\ref{eq:P}),
where 
\begin{align}
F( N_{\rm B} , N_{\bar{\rm B}} ) 
= P_{\langle N_{\rm B} \rangle}( N_{\rm B} )
P_{\langle N_{\bar{\rm B}} \rangle}( N_{\bar{\rm B}} ). 
\end{align}

In this Rapid Communication, 
we derived relations between the baryon and proton number cumulants, 
Eqs.~(\ref{eq:Np})$-$(\ref{eq:Np4}) and
(\ref{eq:NB})$-$(\ref{eq:NB4}), respectively, 
on the basis of the binomial 
nature of (anti-)nucleon isospin numbers in the final state.
These results enable to immediately determine the baryon number 
cumulants with experimental results in heavy ion collisions, 
which will provide significant information about the QCD phase 
diagram.
Though these results are obtained for the isosymmetric case,
incorporation of nonzero isospin density is 
straightforward and will be discussed elsewhere.

The authors are thankful for stimulating discussions at the workshop
``Frontiers of QCD Matter'' held at Nagoya University, Japan, 
on 8 June 2011, and thank H.~Torii for organizing the event.
This work is supported in part by Grants-in-Aid for 
Scientific Research by Monbu-Kagakusyo of Japan 
(No.~21740182 and No.~23540307).

\end{document}